# Scheil-Gulliver simulations for the design of functionally graded alloys by additive manufacturing using pycalphad

Brandon Bocklund[1*], Lourdes D. Bobbio[1], Richard A. Otis[2], Allison M. Beese[1,3], Zi-Kui Liu[1]

[1]Department of Material Science and Engineering, Pennsylvania State University, University Park, PA 16802

[2]Engineering and Science Directorate, Jet Propulsion Laboratory, California Institute of Technology, Pasadena, CA 91109

[3]Department of Mechanical Engineering, Pennsylvania State University, University Park, PA 16802

* Corresponding author, email: bjb54@psu.edu

**Abstract:** Additive manufacturing (AM), through directed energy deposition, supports planned composition changes between locations within a single component, allowing for functionally graded materials (FGMs) to be developed and fabricated. The formation of deleterious phases along a particular composition path can cause significant cracking during the AM build process that makes the composition path unviable to produce these FGMs, but it is challenging to predict which phases will be present in as-built additively manufactured parts by analyzing only equilibrium phase relations. Solute segregation during solidification can lead to the formation of non-equilibrium phases that are stable at compositions far from the nominal composition of the melt, leading to crack formation. In this work, we developed a Scheil-Gulliver simulation tool based on pycalphad. We used this tool to compare the non-equilibrium phases predicted to form during solidification using the Scheil-Gulliver model with experimentally measured phases at several locations with different composition in a Ti-6Al-4V to Invar-36 FGM and a commercially pure Ti to Invar-36 FGM. We showed that the phases predicted to form by the Scheil-Gulliver model outperform the predictions made by assuming equilibrium solidification. Further, we demonstrated the use of our Scheil-Gulliver simulation tool as a method of screening potential FGM pathways by calculating the solidification phase fractions along the experimental gradient path in composition space.

1. **Introduction**

In functionally graded materials (FGMs), phases and properties are tailored spatially within a single component through changes in composition. Directed energy deposition (DED) additive manufacturing (AM) is a method for building parts layer-by-layer by feeding powder from a nozzle into a melt pool produced by a laser [1]. The compositional control afforded by DED is difficult or impossible to achieve with traditional metallurgical processing techniques and can be used to produce FGMs by incrementally varying the composition of the deposited alloy during the build process by changing the volume of powder feedstocks deposited into the melt pool from multiple nozzles.

An FGM with terminal alloys of a Ti-based alloy and Invar-36 (Fe-36Ni wt%, Invar) can take advantage of the high strength to weight ratio of Ti [2] with, ideally, a smooth transition to the low thermal expansion of Invar [3]. Ti and Invar cannot be directly joined by a linear grading due to the formation of deleterious Fe-Ti intermetallic compounds such as $Fe_2Ti$, C14-type Laves phase, and B2 (Fe,Ni)Ti, as solidification products when directly mixing these alloys in liquid state joining, all of which have been found in both welds [4,5] and FGMs [6,7]. A predictive method for FGM design that can determine the phases that may form during the build process, in the spirit of the Integrated Computational Materials Engineering (ICME) approach, is desirable.

Previously, researchers have predicted FGM solidification products using equilibrium calculations at isothermal temperatures along the solidification path [6,8,9]. The representative isothermal temperatures are chosen by matching the computationally predicted phase fractions at that temperature which agrees well with the phase fractions found experimentally. The solidus temperature across a planned composition path could be used as a guide for choosing one or more representative isotherms, but this approach is limiting when considering new composition paths or paths where the solidus temperature changes significantly. In practice, representative isothermal analysis may be able to capture the frozen-in phase fractions and compositions, but cannot describe the phase transformations that take place throughout the build process.

Currently, methods are being developed that aim to predict viable composition paths for FGMs. A path planning algorithm proposed by Kirk *et al.* [10] finds optimal paths between two points in composition space within a range of candidate temperatures, where an optimal path may minimize the path length, maximize the distance from deleterious phases, or use another objective. This type of path planning algorithm relies on designing optimal compositional paths for FGMs by considering solid state reactions over a range of representative temperatures via equilibrium calculations using thermodynamic databases. While path planning approaches of this type are more predictive and transferable than choosing a single isotherm, there are several factors preventing widespread adoption of these methods. One drawback of these algorithms is that they cannot easily consider phases that are metastable or phases that form during solidification and thermal cycling. Furthermore, mapping the obstacles for multicomponent alloys can be computationally intensive because the dimension of the space where a path may exist increases with the number of elements or components, which may limit the capability for path planning algorithms as multicomponent path screening tool.

The rapid solidification in additive manufacturing leads to solute segregation and formation of phases that are not in the equilibrium solidification path. Predicting the actual solidification path is critical for alloys made by AM. In multicomponent FGMs, the compositions on the gradient path can form a variety of phases through solidification by solute segregation; therefore, an efficient computational screening approach is desired. The Scheil-Gulliver solidification model [11,12] can predict the phases and compositions that solidify from a melt using only a thermodynamic description of the phases in a system. The Scheil-Gulliver model has been widely applied in the welding literature [13], but it is not typically used in bulk processes such as casting due to the relatively slower cooling rates. Laser AM is often concerned with building bulk parts through material deposition processes similar to welding, so the rapid cooling effects captured by the Scheil-Gulliver model should be considered. The key assumptions in the Scheil-Gulliver model are that the liquid phase is well-mixed and homogenous in composition as the melt solidifies, while the back-diffusion from the formed solid phases is negligible. During solidification in AM, the melt pool is well-mixed due to Marangoni flow [14], justifying the use of the Scheil-Gulliver model. In addition, Keller *et al.* [15] showed good agreement between the solidification paths predicted by the Scheil-Gulliver model and a diffusion simulation that considered the kinetic behavior explicitly for Inconel 625. The good agreement between Scheil-Gulliver and diffusion simulations indicates that the fast cooling in the AM process is approximated well by the Scheil-Gulliver model, even though Inconel 625 contains elements that have high diffusivities in the solid, such as C, which would cause deviation from the Scheil-Gulliver model.

Scheil-Gulliver solidification simulations are a part of many commercial thermodynamics packages based on the CALculation of PHAse Diagrams (CALPHAD) method, such as Thermo-Calc [16] and Pandat [17]. Recently, the open-source pycalphad software has been developed [18] for solving the multi-component, multi-phase Gibbs energy minimization problem within the CALPHAD method, using symbolic representations of Gibbs energy models to provide a more flexible way to develop new Gibbs energy and property models without changing the internal software. As pycalphad becomes more widely used to develop and fit new Gibbs energy models [19,20], an open-source solidification simulation tool that can use any Gibbs energy model supported by pycalphad can broaden the impact of solidification simulations performed with existing and new CALPHAD databases. A Scheil-Gulliver simulation tool based on pycalphad was developed as part of this work with a focus on providing simulation results in a user-friendly data structure that can be post processed and combined into higher level analysis enabling high-throughput simulations in multi-component composition space.

To effectively determine the viability of a feedstock composition for AM, the phases that will form during the solidification of the melt pool should be quantified in terms of the expected phase fractions. The as-solidified phase fractions can be used to design a composition path that does not produce deleterious phases in excess of an acceptable amount to avoid cracking. The two methods to predict phase fractions compared here are equilibrium and Scheil-Gulliver solidification. This work demonstrates the necessity of the Scheil-Gulliver solidification model as a design tool for screening predictions of solidification products and viable FGM compositional paths compared to equilibrium calculations alone. The equilibrium and Scheil-Gulliver solidification behavior of Fe-Ni-Ti alloys corresponding to representative layers in a commercially pure Ti to Invar-36 (CP Ti/Invar) FGM and Ti-6Al-4V to Invar-36 (Ti-6Al-4V/Invar) FGM are predicted using a thermodynamic description of the Fe-Ni-Ti system within the CALPHAD method at compositions measured by energy dispersive X-ray spectroscopy (EDS). The predictions are validated by electron backscatter diffraction (EBSD) phase characterization of selected regions of both FGMs. Using this approach, we show that Scheil-Gulliver solidification can better predict the phases that form during solidification in AM.

2. **Scheil-Gulliver Model**

The Scheil-Gulliver solidification model [11,12] predicts the solid phases that precipitate from a melt that changes in composition due to the segregation of solute species from the liquid through local equilibrium at the solid/liquid interface. Scheil-Gulliver solidification considers the fast cooling case where the mass transport in the liquid is fast enough for perfect mixing in the liquid, but the diffusivity in the solid is low enough so that there is no diffusion in the solid. These conditions give an upper limit for the partitioning of mass between solid and liquid. These assumptions lead to the solid depleting one or more constituents from the liquid phase as the temperature of the melt decreases during the simulation, ultimately ending a at a eutectic point where the remaining liquid solidifies into the eutectic phases. In AM, liquid diffusion and Marangoni flow contribute to a well-mixed liquid throughout solidification.

The assumptions of local equilibrium, perfect liquid mixing, and no solid diffusion allows Scheil-Gulliver simulations to make time-independent predictions of the formation of solid phases from liquid using only the thermodynamic description of a system given the following algorithm [21]:

1) Given the current phase fraction of liquid, $f_{\text{liquid},i}$, in the system at timestep $i$ (initially $f_{\text{liquid},0} = 1$), perform an equilibrium calculation at the given conditions: current temperature, $T_i$, fixed pressure, and composition of the system (initially the overall composition of the alloy).

2) Partition the current fraction of liquid into liquid and solid based on the phase fractions from the equilibrium calculation, $f_{\text{liquid},i+1} = f_{\text{liquid},i}\, \phi_{\text{liquid}}$, where $\phi_{\text{liquid}}$ is the phase fraction of liquid in the equilibrium calculation. Store the solid phases, amounts, and their compositions for post processing.

3) Reduce the temperature by $T_{i+1} = T_i - \Delta T$, where $\Delta T$ is temperature step size, and set the composition of the system to the new composition of the liquid phase from the equilibrium calculation.

4) Repeat steps 1-3 until $f_{\text{liquid},i}$ is below a user-defined threshold or a eutectic is reached.

We have written an open-source Python package called "*scheil*" that implements this approach using pycalphad as the thermodynamic calculation engine to enable the flexible use of arbitrary Gibbs energy models for the liquid and solid phases. The *scheil* software is distributed on the Python Package Index (PyPI) [22]. The software has been designed to simulate both equilibrium solidification and Scheil-Gulliver solidification. Simulation results for both Scheil-Gulliver and equilibrium solidification are stored in a *SolidificationResult* data structure that provides access to the phase fractions and phase compositions of all the phases in the system throughout the simulation. During the simulation, additional candidate grid points corresponding to the site fractions of the equilibrium phases found at a particular temperature are adaptively added to the point grid used in pycalphad for starting point generation and global minimization [23]. Since the site fractions of the stable phases at $T_i$ are likely close to those at $T_{i+1}$ both the performance and accuracy of the energy minimization in pycalphad are improved by starting near the global minimum solution.

A key feature of the *scheil* software developed in this work is the ability to distinguish and treat separately ordered and disordered configurations of phases that are modeled using a partitioned order-disorder model [24,25] in a way that is transparent to the user. The partitioned order-disorder model is commonly used to describe B2 ordering in bcc alloys and $L1_2$ and $L1_0$ ordering in fcc alloys using a single phase to describe the Gibbs energy of both ordered and disordered configurations. Disordered configurations occur when the site fractions for any species is equal in all sublattices. This feature allows the solidified phase fractions of the ordered and disordered configurations to accumulate separately, even if those phases both form at the same temperature step, which is an important distinguishing feature compared to commercial implementations, such as the SCHEIL or POLY3 modules in Thermo-Calc, the latter of which can distinguish order and disordered configurations of order-disorder partitioned phases but it is difficult track and store which configuration is stable throughout the simulation so the phases can be treated separately during post processing. The *scheil* software is designed for high-throughput screening and is capable of performing multiple simulations by looping over a series of compositions that may correspond to a linear or non-linear gradient path, or a grid in multi-component composition space as part of a more complex data-driven path planning simulation.

In this work, a thermodynamic description for Fe-Ni-Ti modeled over the entire composition range within the CALPHAD method by De Keyzer et al. [9] was used. The CALPHAD method uses a description of the Gibbs energy of each phase, partitioned into the contributions from the surface of reference, configurational entropy, physical models (e.g., magnetism), and excess contributions [26]. The CP Ti/Invar FGM contains only Fe, Ni, and Ti, so the compositions measured experimentally by EDS were used directly in the computations. The Al and V measured in the Ti-6Al-4V/Invar FGM were neglected in the computational simulations and the remaining measured compositions were normalized to the Fe-Ni-Ti ternary system. The phase fractions of each layer studied in both CP Ti/Invar and Ti-6Al-4V/Invar FGMs are compared to the equilibrium and Scheil-Gulliver solidification products. Equilibrium solidification products are defined as the equilibrium phase fractions at the solidus line at the overall composition of the layer of interest. Scheil-Gulliver solidification products are defined as the cumulative solid phase fractions at the end of the solidification simulation, starting with the same overall layer composition.

Scheil-Gulliver simulations were run starting from the measured overall composition above the liquidus temperature with a step size of 10˚C until the fraction of material solidified reached 0.9999. The complex Gibbs energy surfaces for the ternary sublattice models in the Fe-Ni-Ti system, such as the C14 phase that uses three sublattices occupied by all three elements, required the use of adaptively sampling the convex hull of the energy surface of each phase to add extra low energy composition sets for the pycalphad global minimization. It is expected that this treatment would have been required in Thermo-Calc as well, since for certain compositions the Scheil-Gulliver simulation performed within the SCHEIL module would terminate before reaching a eutectic point. All the results in this publication were created using scheil version 0.1.2 [27] and pycalphad version 0.8.1 [18]. A Jupyter notebook containing all of the code to reproduce the results in this publication may be found in the supplemental materials [28].

## 3. Experimental Methods

The CP Ti/Invar and Ti-6Al-4V/Invar FGM samples were fabricated using a directed energy deposition system (RPM 557 Laser Deposition System) with a YAG laser in an argon atmosphere in order to prevent oxidation. The RPM 557 Laser Deposition System can deposit varying mixtures of feedstock powders during fabrication, allowing for the change in volume fraction of powder as a function of position, as required for FGM fabrication. The samples were deposited in 75 layers with a hatch spacing of 0.58 mm and a layer height of 0.38 mm to form posts with a 15 mm square base and a height of 28.5 mm. The CP Ti/Invar FGM was fabricated using a laser power of 800 W and a powder mass flow rate ranging from $6.7 \times 10^{-4}$ g/s to $5.8 \times 10^{-1}$ g/s, while the Ti-6Al-4V/Invar FGM was fabricated using a laser power of 900 W and a powder mass flow rate ranging from $6.6 \times 10^{-4}$ g/s to $5.8 \times 10^{-1}$ g/s. Both were operated at a laser scanning speed of 12.7 mm/s.

Ti-6Al-4V and Invar powder was used for the Ti-6Al-4V/Invar sample with powder diameters ranging from 45 μm to 177 μm (-80/+325 mesh size) for both powders. The CP Ti/Invar FGM was fabricated using the Invar powder and CP Ti powder with diameters ranging from 45 μm to 150 μm (-100/+325 mesh size). A schematic of the additively manufactured FGMs is shown in **Figure 1**. For both the CP Ti/Invar and Ti-6Al-4V/Invar FGMs, initially 21 layers of 100 vol% CP Ti and 100 vol% Ti-6Al-4V, respectively, were deposited. In the gradient regions, the amount of the starting powder (i.e., CP Ti or Ti-6Al-4V) was decreased by 3 vol% per layer, and replaced by 3 vol% of Invar per layer for 32 layers until a composition of 100 vol% Invar was achieved. Finally, 22 layers of 100 vol% Invar were deposited. The as-built samples were removed from the baseplate using wire electrical discharge machining and sectioned vertically to expose the cross-section of the sample. The sectioned samples were mounted in epoxy and prepared for analysis using standard metallographic techniques, with a final polish using 0.05 μm silica suspension.

The chemical composition was measured and the phases were identified within the gradient regions of both samples. A scanning electron microscope (SEM, FEI Quanta 200) with an attached silicon drift detector (Oxford X-act PentaFET Precision) was used to for EDS measurements. Phase identification was performed using EBSD (Oxford Nordlys Max2).

4. **Results and discussion**

4.1 **Fe-Ni-Ti liquidus projection**

**Figure 2** shows the calculated Fe-Ni-Ti liquidus projection from the thermodynamic database assessed by De Keyzer [29]. The liquidus projection shows the first solid phase to form from the liquid upon cooling. The black solid lines in the diagram indicate monovariant phase equilibria, where two or more solid phases are in equilibrium with the liquid phase, projected along the temperature axis. The monovariant lines coalesce into invariant phase equilibria of three solid phases and the liquid phase, being peritectic or eutectic. A Scheil-Gulliver solidification simulation passes peritectic invariant reactions to lower temperature due to the remaining liquid and always ends at a eutectic invariant reaction point. The arrows on the monovariant lines point in the direction of decreasing temperature such that following the arrows down the monovariant lines will ultimately lead to a eutectic reaction. In the Fe-Ni-Ti system there are three eutectic points, which are labeled in **Figure 2** as E1, E2 and E3. The E1 eutectic point, ending in the Ni-Ti binary, has liquid, bcc, and $NiTi_2$ in equilibrium, E2 has liquid, Laves C14, B2, and $Ni_3Ti$, and E3 has liquid, Laves C14, fcc, and $Ni_3Ti$. The colored lines depict the solidification paths from the selected compositions and the corresponding eutectic point for that composition. The Laves C14 phase has the highest melting point in the Fe-Ni-Ti system, with a maximum melting point of 1460°C near the composition forming $(Fe, Ni)_2Ti$ with equal amounts of Fe and Ni. Alloy compositions near this composition of maximum melting temperature can reach any of the three eutectic points, depending on how the composition deviates from the maximum. The dashed lines in the Laves C14 region indicate the regions of composition space that will lead to the different eutectic points. The B2 region has a ridge of local maxima in the melting point, shown as a dashed line in **Figure 2**, such that compositions on the Ni-rich side will take solidifications paths ending at E2, while compositions on the Ni-poor side of this line, will take solidification paths ending at eutectic E1.

**4.2 Comparison of Scheil-Gulliver predictions with EBSD results**

Two layers within the CP Ti/Invar FGM were analyzed (layer 24 and layer 32) and four layers within the Ti-6Al-4V/Invar FGM were analyzed (layers 26, 33, 35, and 45). **Table 1** shows the overall compositions of these six layers as measured by EDS. The EBSD phase maps for the analyzed layers in both FGMs are shown in **Figure 3**. Note that the disordered bcc and ordered B2 phases cannot be distinguished by EBSD because they share the same crystallographic parameters. While the ordered B2 has allowed reflections that are forbidden in the disordered bcc, the corresponding Kikuchi bands overlap, resulting in identical Kikuchi patterns that cannot be differentiated when indexing the patterns. Therefore, the phase regions corresponding to the disordered bcc phase, where Fe, Ni and Ti are distributed randomly on all crystallographic sites, and the ordered B2 phase, where Fe, Ni, and Ti are found preferentially on centers or corners of the bcc lattice sites, will be referred to as B2/bcc. Tables 2-7 show the measured EBSD phase fractions in each layer compared with the computationally predicted phase fractions. The "EBSD" column includes the raw EBSD phase fraction data and the fraction of the region that was unidentified, which is the area fraction of the scanned regions where the Kikuchi patterns could not be resolved, and thus this area was not identified to be any specific phase by the analysis software. In the "EBSD normalized" column, the phase fractions of the identified phases are normalized to remove the unidentified area for comparison with the computed phase fractions. Tables 2-7 compare these normalized phase fractions to the phase fractions predicted by the equilibrium calculations and by the Scheil-Gulliver solidification simulations. Note that for the computational simulations, the bcc and B2 rows are shown combined on the left and separated on the right. The separated bcc and B2 values are shown since these phases are easily distinguished computationally and would exhibit different, potentially undesired properties, but they are also shown combined for easier comparison to the experimental results.

The Scheil-Gulliver model predicted that layer 24 (91 vol% Ti, 9 vol% Invar) of the CP Ti/Invar FGM (**Table 2**) should follow the path to bcc then to eutectic E1 with bcc/NiTi$_2$. Layer 32 (67 vol% Ti, 33 vol% Invar) of the CP Ti/Invar FGM and layers 26 (85 vol% Ti-64, 15 vol% Invar) and 33 (64 vol% Ti-64, 36 vol% Invar) of the Ti-6Al-4V/Invar FGM (**Table 3**, **Table 4** and **Table 5**, respectively) were all predicted to go towards the B2 monovariant line, then down to the bcc/NiTi$_2$ eutectic E1, with layer 33 containing the Laves C14 phase based on the starting composition, crossing the monovariant line corresponding to the peritectic reaction involving liquid, Laves C14, and B2. In the Ti-6Al-4V/Invar FGM, layer 35 (58 vol% Ti-64, 42 vol% Invar) and layer 45 (27 vol% Ti-64, 73 vol% Invar) start in the Laves C14 and fcc regions, respectively, and go to the Laves C14/fcc monovariant line until reaching the Laves C14/fcc/Ni$_3$Ti eutectic, E3. The main phases present are correctly predicted by both equilibrium and Scheil-Gulliver models, but there are some discrepancies between the predicted and experimental phase fractions that will be addressed in the next section. However, the phase fractions predicted by Scheil-Gulliver solidification are significantly closer in layer 26 in the Ti-6Al-4V/Invar FGM. Scheil-Gulliver solidification products for layer 35 of the Ti-6Al-4V/Invar FGM (and **Table 6**) compare more favorably to the experimental compositions than the equilibrium solidification products, which are predicted to contain almost no Ni$_3$Ti.

In the CP Ti/Invar FGM, layer 24 (**Table 2**) the microstructure shows that the alloy initially crystallizes into B2, covering 69.9% of the analyzed area, followed by the NiTi$_2$ phase covering 28.2%. In layer 24 of the CP Ti/Invar FGM, Scheil-Gulliver solidification predicts predict the formation of the B2, bcc and NiTi$_2$ phase, with phase fractions in good agreement with experimental results. The equilibrium solidification calculation predicts that only bcc forms, not NiTi$_2$. Scheil-Gulliver simulations also predict the formation of more phases in the Ti-6Al-4V/Invar FGM for layers 33 (**Table 5**) and 45 (**Table 7**); however, layer 33 contains a small amount of fcc in the experiment while the Scheil-Gulliver simulation predicts Ni$_3$Ti. The detection of a new phase in the Scheil-Gulliver simulation is a significant improvement over the equilibrium calculation alone because it would suggest that even this relatively moderate introduction of Fe and Ni into Ti leads to the formation of undesired phases. This demonstrates they key value of the Scheil-Gulliver solidification over equilibrium solidification, since the compositions where new phases form may not be accessed without the solute partitioning that occurs in the Scheil-Gulliver model.

For layer 32 in the CP Ti/Invar FGM (**Table 3**), the equilibrium calculation predicts that 70.8% B2 with the balance NiTi$_2$, while the Scheil-Gulliver solidification simulation predicts that B2 will primarily form (67.8%), but will also form bcc and NiTi$_2$ as the liquid is enriched with Ni and Ti towards eutectic E1, with 8.8% bcc and 23.6% NiTi$_2$. For the EBSD results, the normalized phase fractions give an almost 1:1 ratio between bcc/B2 and NiTi$_2$; however, 17.1% was unidentified, which could contribute to the discrepancy between the computational predictions. Large total unidentified areas are also found in Ti-6Al-4V layers 26 (**Table 4**) and 35 (**Table 6**), with 11.3% and 12.2%, respectively. In layer 26, there is significant disagreement between the experimental phase fractions and the predicted equilibrium and Scheil-Gulliver phase fractions, but the Scheil-Gulliver simulations are closer to the experimental phase fractions than the equilibrium calculation. Similar to CP Ti layer 32, the Scheil-Gulliver simulation predicts the formation of both bcc, B2 and NiTi$_2$ while the equilibrium simulation predicts that only B2 and NiTi$_2$ will be solidification products. In layer 35, significant amounts of several phases are present, and the unidentified regions do not appear localized to any particular phase. Both equilibrium and Scheil-Gulliver calculations predict the presence of Laves C14, $Ni_3Ti$ and fcc phases, but equilibrium predicts a phase fraction of only 0.3% $Ni_3Ti$, while Scheil-Gulliver simulation predicts 5.9% $Ni_3Ti$ compared to the experimental value of 19.2% $Ni_3Ti$. The 12.2% of unidentified phase amounts could account for the discrepancy in the normalized phase fractions, depending on how it is distributed across the phases in reality.

**4.3 Discussion**

The phases present in as-built parts must form through either solidification or subsequent solid state phase transformations. Here it was shown that the phases found at different compositions along linear paths between Ti and Invar are well matched to the phases predicted to form in a highly segregating solidification process within the assumptions of the Scheil-Gulliver model, while equilibrium solidification, the lower bound for segregation, fails to predict the formation of experimentally present phases. Approaching the tailoring of FGMs through computational solidification simulations offers a clear advantage to the more common approach of selecting representative temperatures for thermodynamic analysis because it can be predictively applied to new gradient paths or new materials systems to guide the experimental work. However, there are still some outstanding discrepancies between the solidification predicted phases and the experimentally determined phases that should be reconciled. For example, small amounts of hcp solution were detected in several layers, which could not be predicted to form in a solidification scheme because the only stable hcp phase in the Fe-Ni-Ti system is the low temperature Ti-rich hcp phase, therefore a solid state transformation must have occurred. This has also been observed in other material systems including the formation of the Fe-(Cr,V) σ phase that has been observed in several FGMs [8,9] despite not being a primary solidification product in those alloy systems.

Since Scheil-Gulliver solidification is an upper bound for segregation and phase formation along the solidification path, the true phases and compositions that are found in the as-solidified alloy must be between the equilibrium solidification and Scheil-Gulliver solidification. After the locally melted material has completely solidified, the heat from the recently solidified material must be dissipated, primarily through conduction to layers and baseplate below, with nucleation and growth kinetics controlling the solid-state phase transformations from the solidification to final alloy products. In addition, a challenge for using the Scheil-Gulliver model for path planning design is that re-melting of previously deposited layers in FGMs can lead to melt pool compositions that are different from the composition of the planned stock material at that layer, which is further complicated because the re-melted material may be the solute-rich eutectic that was last to solidify in the previous layer. The comparisons in this work were not affected by any re-melting because the compositions used in the computational analysis were the measured EDS composition for each layer, rather than the planned composition.

Liquidus projections show the primary crystallization phases during solidification. When combined with isothermal contour lines, as shown for Fe-Ni-Ti by De Keyzer *et al.* [29], the liquidus projection maps the liquidus surface across the composition shown in the diagram. Since Scheil-Gulliver solidification follows the composition of the liquidus, the liquidus projection is qualitative map of the solidification path of any alloy as it solidifies under Scheil-Gulliver solidification. Liquidus projections, which generalize to multicomponent systems, can be used as a design tool to screen for potential gradient paths. In Fe-Ni-Ti alloys, it is known that $NiTi_2$, and Laves C14 phases are brittle compounds [29] that may act as crack nucleation sites. The liquidus projection in **Figure 2** shows clearly that no smooth transition could be made between pure Ti and the bcc or fcc Fe/Ni-rich region without passing through a deleterious phase. Abrupt transitions should not be used in additively manufactured FGMs because the compositions leading to deleterious phase formation will still be accessed through either solidification segregation or solid-state diffusion; therefore, any transitions from pure Ti to Fe, Ni, or an Fe-Ni alloy are not possible using only the elements in this ternary system. In this way, liquidus projections for ternary and multicomponent systems could be used as a visual tool for selecting candidate composition paths or regions out of more computationally complex FGM path simulations by only selecting paths where the primary crystallization product is a favorable phase.

The *scheil* software can be used to simulate Scheil-Gulliver solidification along an arbitrary composition path, which are equivalent to the Scheil ternary projection (STeP) diagrams introduced recently by Moustafa *et al.* [30]. It can therefore be used as a quantitative, data-driven tool to screen all of composition space or a subspace. An example of using this design tool is presented in **Figure 4** for the linear gradient between Ti and Invar in this work. **Figure 4a** shows the as-solidified phase fractions from the Scheil-Gulliver simulation along the composition path and **Figure 4b** shows the as-solidified phase fractions under equilibrium solidification conditions. Both figures have deconvoluted the disordered bcc phase from the ordered B2 phase, which are modeled as the same phase in the thermodynamic database from De Keyzer *et al*. [29]. In the Scheil-Gulliver solidification simulations, as soon as enough Invar is added to Ti along the gradient path to make B2 the primary crystallization phase, at least three phases are predicted to form during solidification at any particular composition through the rest of the gradient until pure Invar is reached. Nowhere along the path in any of the gradient region is only one phase present, in agreement with the experimental analysis. The phase fractions along the gradient path under equilibrium solidification conditions show that the Ti-rich and Invar-rich compositions form single phase bcc and fcc, respectively. The intermediate equilibrium solidification simulations never predict more than two solid phases to form at any composition and the $Ni_3Ti$ is never predicted to form. This demonstrates the power of using the Scheil-Gulliver model to predict the solidification phases during the AM process, since several regions along the path would be predicted to form only one phase using equilibrium calculations alone. Furthermore, since cracks may nucleate from the formation of deleterious phases [31], Scheil-Gulliver simulations could be used to predict upper limit for the formation of deleterious phases that nucleate cracks, enabling the design space for traditional AM alloys to be explored within a tolerance for the formation of crack-nucleating phases. In this way, CALPHAD modeling can be used to predict which phases can form in the additive manufacturing process.

5. **Conclusion**

The compositions and phase fractions of two FGMs, one from CP Ti/Invar and one from Ti-6Al-4V/Invar were characterized computationally and experimentally. Experimental EBSD phase maps of selected representative layers of the two FGMs showed the various phases present as a function of FGM composition. The experimental phase fraction data were compared to the phases that were predicted to form by equilibrium solidification and Scheil-Gulliver solidification within the CALPHAD method. The phases predicted by Scheil-Gulliver simulations, representing the upper bound of solute partitioning between the liquid and solid, were able to better predict the phases formed in the as-built FGMs, demonstrating that the rapid cooling and melt behavior in the additive manufacturing process are well approximated by the Scheil-Gulliver solidification model. Liquidus projections and Scheil-Gulliver simulations can be used as qualitative and quantitative tools within an ICME framework for assessing the viability of FGMs by predicting solidification products in complex multicomponent FGMs with non-linear composition paths, serving as a guide to the rapid prototyping capabilities of additive manufacturing.

**Acknowledgments**

BB and ZKL were supported by a NASA Space Technology Research Fellowship (Grant 80NSSC18K1168). LDB was supported by an NDSEG Fellowship. This work was also partially supported by the NASA Jet Propulsion Lab (Grant JPL 1596437) and the Office of Naval Research (Grant N00014-17-1-2567). Part of the research was carried out at the Jet Propulsion Laboratory, California Institute of Technology, under a contract with the National Aeronautics and Space Administration (80NM0018D004).

**Figures**

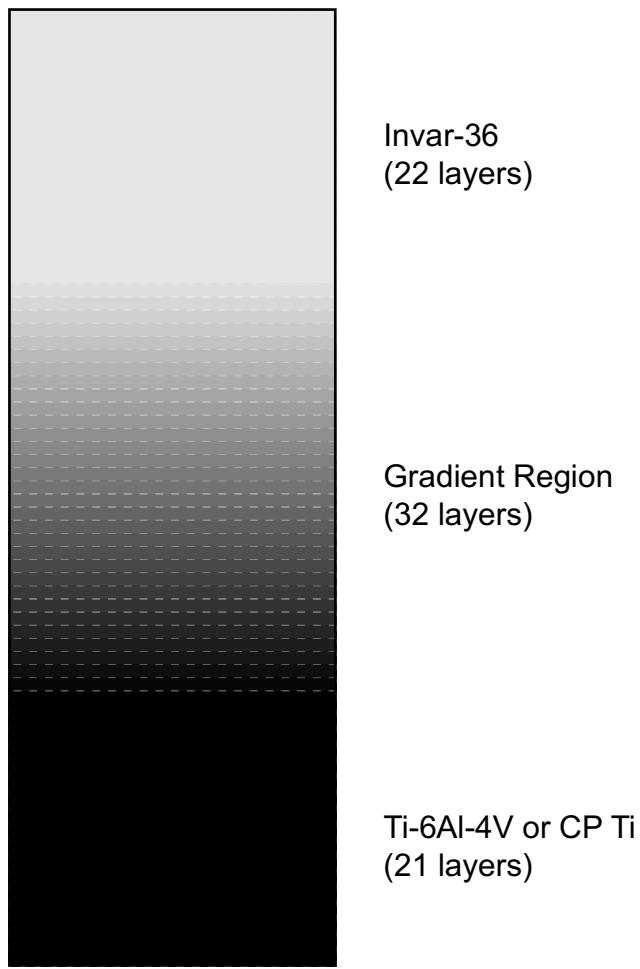

**Figure 1.** Schematic of the planned gradient path for the Ti-6Al-4V/Invar and CP Ti/Invar FGMs. The FGMs were deposited as posts with square bases that were 15 mm in length. The height of each layer was 0.38 mm for a total height of approximately 28.5 mm. In both samples, cracking occurred in the gradient regions.

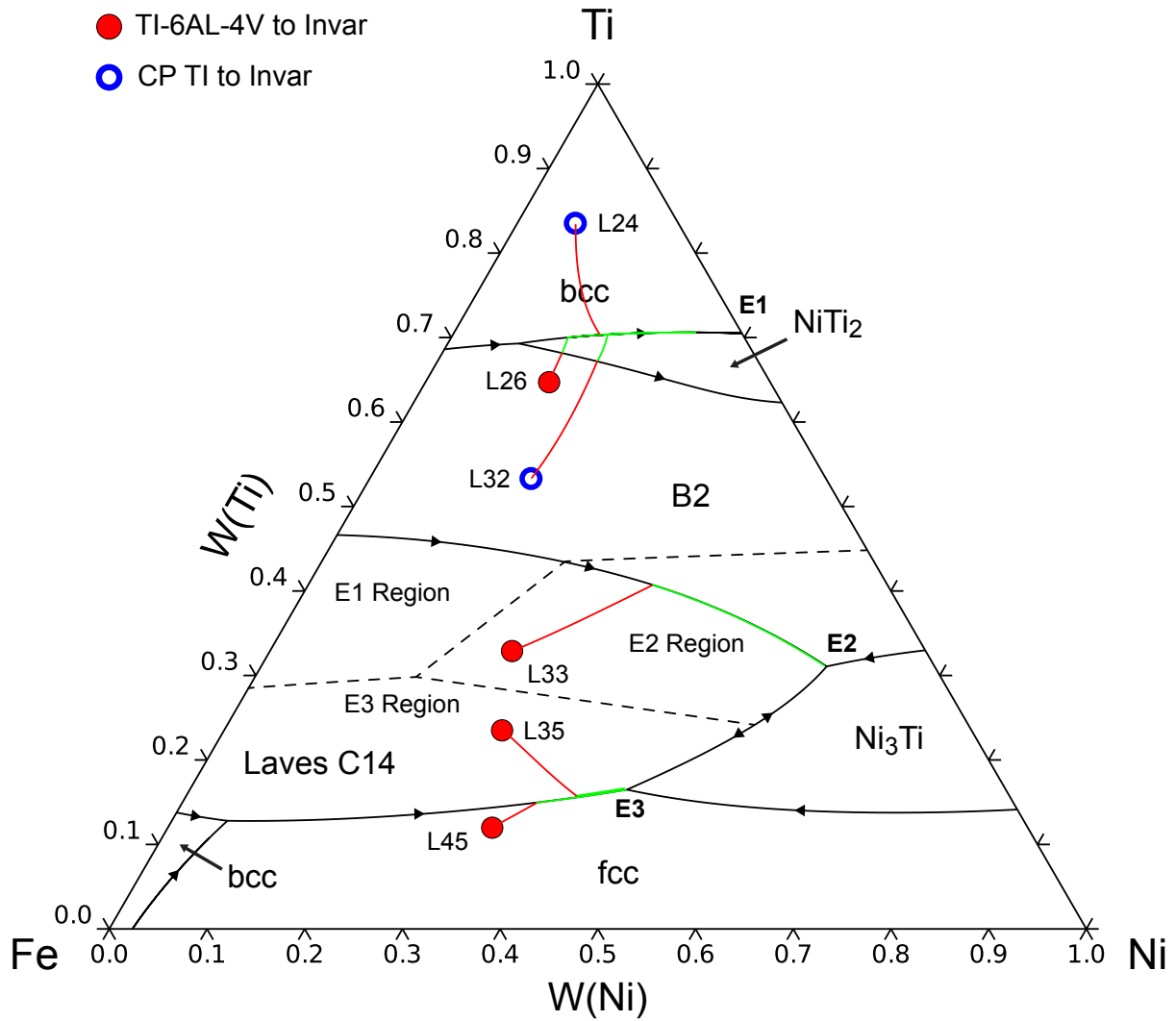

**Figure 2.** Liquidus projection of the Fe-Ni-Ti system based on the modeling by De Keyzer et al. [29]. The compositions studied in this work are marked by layer number and colored in red (closed circles) and blue (open circles) for the Ti-6Al-4V and CP Ti samples, respectively. The dashed lines separating different regions where the Laves phase is the primary crystalline product correspond to the eutectic at which the composition in that region will end up. The regions and corresponding eutectics are labeled **E1**, **E2**, and **E3**.

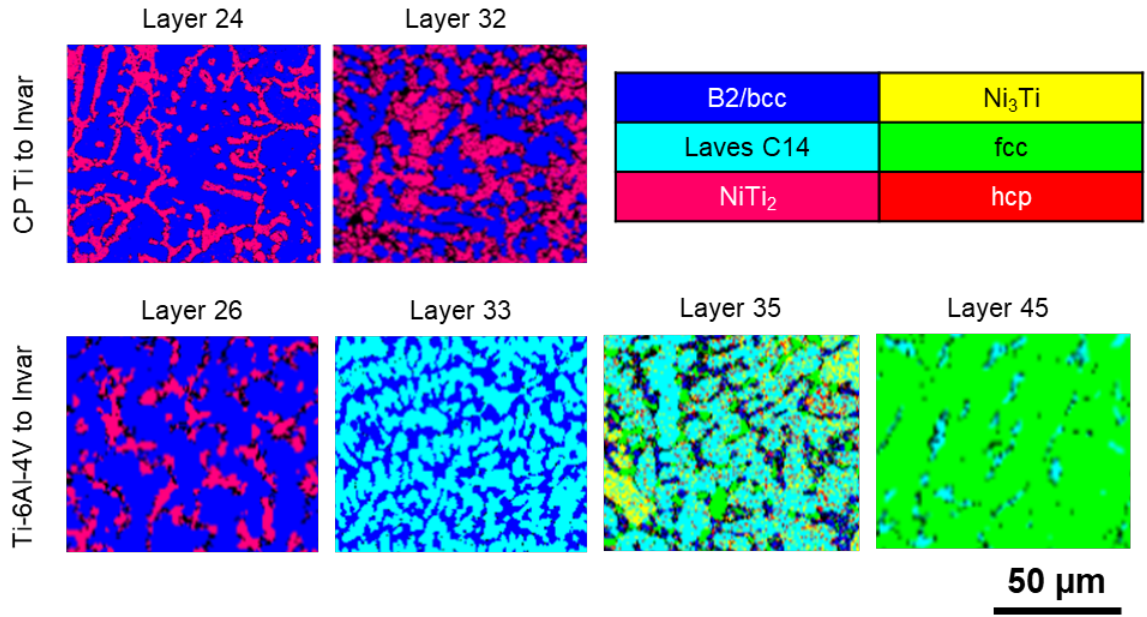

**Figure 3.** EBSD phase maps of layers (a) 24 (91 vol% Ti, 9 vol% Invar) and (b) 32 (67 vol% Ti, 33 vol% Invar) of the CP Ti/Invar FGM and layers (c) 26 (85 vol% Ti-64, 15 vol% Invar), (d) 33 (64 vol% Ti-64, 36 vol% Invar), (e) 35 (58 vol% Ti-64, 42 vol% Invar), and (f) 45 (27 vol% Ti-64, 73 vol% Invar) of the Ti-6Al-4V/Invar FGM.

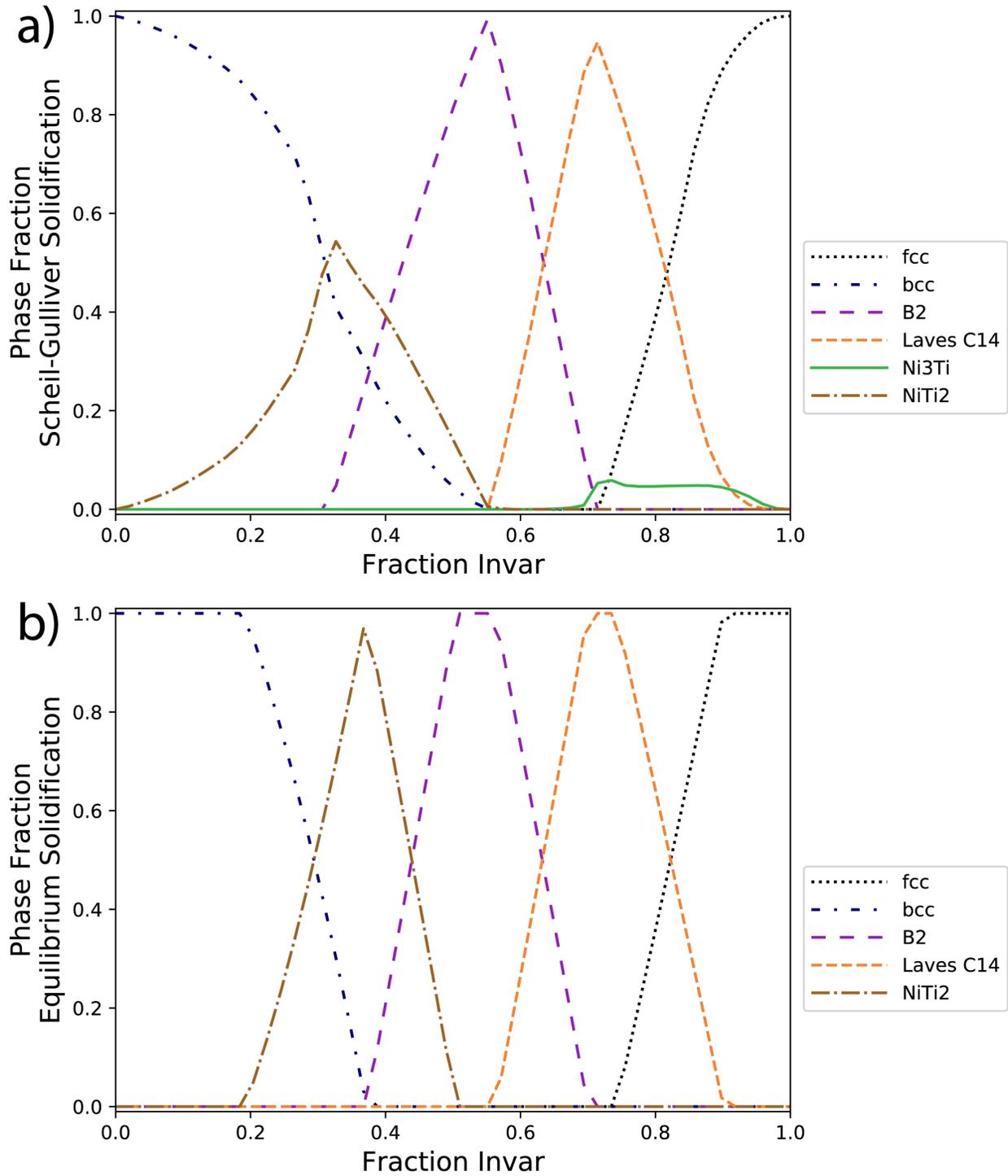

**Figure 4.** Phase fractions of the as-solidified phases predicted along the linear gradient path from Ti to Invar using the Scheil-Gulliver model (a) and the equilibrium solidification (b).

**Tables**

**Table 1.** Overall compositions in weight % for each layer of the CP Ti/Invar and Ti-6Al-4V/Invar FGMs measured by EDS.

|  | Layer | Fe | Ni | Ti | Al | V |
|---|---|---|---|---|---|---|
| CP Ti/Invar | 24 | 10.6 | 6.0 | 83.4 | - | - |
|  | 32 | 30.2 | 16.5 | 53.3 | - | - |
| Ti-6Al-4V/Invar | 26 | 21.1 | 11.8 | 60.3 | 4.1 | 2.7 |
|  | 33 | 40.7 | 24.0 | 31.6 | 2.2 | 1.5 |
|  | 35 | 46.7 | 27.7 | 22.9 | 1.6 | 1.1 |
|  | 45 | 54.0 | 32.7 | 11.8 | 0.9 | 0.7 |

**Table 2.** Experimental phase fractions measured by EBSD and computed phase fractions of the solidification products for layer 24 (Fe-6.0Ni-83.4Ti wt%) of the CP Ti/Invar FGM. The computed phase fractions used the overall composition as measured by EDS.

|  | EBSD | EBSD normalized | Equilibrium solidification | | Scheil-Gulliver solidification | |
|---|---|---|---|---|---|---|
| bcc | 0.670 | 0.699 | 1.000 | - | 0.893 | 0.893 |
| B2 |  |  |  | - |  | - |
| NiTi$_2$ | 0.245 | 0.282 | - | | 0.107 | |
| Laves C14 | 0.025 | - | - | | - | |
| Ni$_3$Ti | - | - | - | | - | |
| fcc | 0.005 | 0.005 | - | | - | |
| hcp | 0.014 | 0.015 | - | | - | |
| unidentified | 0.041 | - | - | | - | |

**Table 3.** Experimental phase fractions measured by EBSD and computed phase fractions of the solidification products for layer 32 (Fe-16.5Ni-53.3Ti wt%) of the CP Ti/Invar FGM. The computed phase fractions used the overall composition as measured by EDS.

|  | EBSD | EBSD normalized | Equilibrium solidification | | Scheil-Gulliver solidification | |
|---|---|---|---|---|---|---|
| bcc | 0.405 | 0.489 | 0.708 | - | 0.766 | 0.088 |
| B2 |  |  |  | 0.708 |  | 0.678 |
| NiTi$_2$ | 0.329 | 0.483 | 0.292 | | 0.234 | |
| Laves C14 | 0.072 | - | - | | - | |
| Ni$_3$Ti | - | - | - | | - | |
| fcc | 0.014 | 0.017 | - | | - | |
| hcp | 0.010 | 0.012 | - | | - | |
| unidentified | 0.171 | - | - | | - | |

**Table 4.** Experimental phase fractions measured by EBSD and computed phase fractions of the solidification products for layer 26 (Fe-12.7Ni-64.7Ti wt%) of the Ti-6Al-4V/Invar FGM. The computed phase fractions used the overall composition as measured by EDS.

|  | EBSD | EBSD normalized | Equilibrium solidification | | Scheil-Gulliver solidification | |
|---|---|---|---|---|---|---|
| bcc | 0.714 | 0.805 | 0.128 | 0.128 | 0.490 | 0.320 |
| B2 |  |  |  | - |  | 0.170 |
| $NiTi_2$ | 0.146 | 0.166 | 0.872 | | 0.510 | |
| Laves C14 | - | - | - | | - | |
| $Ni_3Ti$ | 0.005 | 0.006 | - | | - | |
| fcc | 0.01 | 0.011 | - | | - | |
| hcp | 0.011 | 0.012 | - | | - | |
| unidentified | 0.113 | - | - | | - | |

**Table 5.** Experimental phase fractions measured by EBSD and computed phase fractions of the solidification products for layer 33 (Fe-24.9Ni-32.8Ti wt%) of the Ti-6Al-4V/Invar FGM. The computed phase fractions used the overall composition as measured by EDS.

|  | EBSD | EBSD normalized | Equilibrium solidification | | Scheil-Gulliver solidification | |
|---|---|---|---|---|---|---|
| bcc | 0.348 | 0.362 | 0.217 | - | 0.247 | - |
| B2 |  |  |  | 0.217 |  | 0.247 |
| $NiTi_2$ | - | - | - | | - | |
| Laves C14 | 0.608 | 0.633 | 0.783 | | 0.750 | |
| $Ni_3Ti$ | - | - | - | | 0.003 | |
| fcc | 0.005 | 0.005 | - | | - | |
| hcp | - | - | - | | - | |
| unidentified | 0.039 | - | - | | - | |

**Table 6.** Experimental phase fractions measured by EBSD and computed phase fractions of the solidification products for layer 35 (Fe-28.5Ni-23.5Ti wt%) of the Ti-6Al-4V/Invar FGM. The computed phase fractions used the overall composition as measured by EDS.

|  | EBSD | EBSD normalized | Equilibrium solidification | Scheil-Gulliver solidification |
|---|---|---|---|---|
| bcc | 0.077 | 0.087 | - | - |
| B2 |  |  | - | - |
| $NiTi_2$ | 0.002 | - | - | - |
| Laves C14 | 0.433 | 0.495 | 0.848 | 0.731 |
| $Ni_3Ti$ | 0.192 | 0.219 | 0.003 | 0.059 |
| fcc | 0.102 | 0.116 | 0.149 | 0.210 |
| hcp | 0.073 | 0.083 | - | - |
| unidentified | 0.122 | - | - | - |

**Table 7.** Experimental phase fractions measured by EBSD and computed phase fractions of the solidification products for layer 45 (Fe-33.2Ni-12.0Ti wt%) of the Ti-6Al-4V/Invar FGM. The computed phase fractions used the overall composition as measured by EDS.

|  | EBSD | EBSD normalized | Equilibrium solidification | Scheil-Gulliver solidification |
|---|---|---|---|---|
| bcc | 0.034 | 0.034 |  |  |
| B2 |  |  | - | - |
| $NiTi_2$ | 0.001 | - | - | - |
| Laves C14 | 0.037 | 0.038 | 0.119 | 0.109 |
| $Ni_3Ti$ | 0.014 | 0.014 | - | 0.059 |
| fcc | 0.899 | 0.901 | 0.881 | 0.832 |
| hcp | 0.013 | 0.013 | - | - |
| unidentified | 0.034 | - | - | - |